\documentclass[prl,twocolumn,superscriptaddress,showpacs]{revtex4-1}
\usepackage{graphicx,amsmath,amssymb}
\usepackage{color}
\usepackage[normalem]{ulem}	
\begin{document}

\title{High-pressure single-crystal neutron scattering study of magnetic and Fe vacancy orders in
(Tl,Rb)$_{\rm 2}$Fe$_{\rm 4}$Se$_{\rm 5}$ superconductor}

\author{Feng~Ye}
\affiliation{Quantum Condensed Matter Division, Oak Ridge National Laboratory, Oak Ridge,
Tennessee 37831, USA }
\affiliation{Department of Physics and Astronomy, University of Kentucky,
Lexington, Kentucky 40506, USA}
\author{Wei~Bao}
\email{wbao@ruc.edu.cn}
\affiliation{Department of Physics, Renmin University of China, Beijing
100872, China}
\author{Songxue~Chi}
\author{Antonio~M.~dos~Santos}
\author{Jamie~J.~Molaison}
\affiliation{Quantum Condensed Matter Division, Oak Ridge National Laboratory, Oak Ridge,
Tennessee 37831, USA }
\author{Minghu~Fang}
\author{Hangdong~Wang}
\author{Qianhui~Mao}
\affiliation{
Department of Physics, Zhejiang University, Hangzhou 310027, China}
\author{Jinchen~Wang}
\author{Juanjuan~Liu}
\author{Jieming~Sheng}
\affiliation{Department of Physics, Renmin University of China, Beijing
100872, China}

\begin{abstract}
The magnetic and iron vacancy orders in superconducting
(Tl,Rb)$_2$Fe$_4$Se$_5$ single-crystals were investigated using a high-pressure neutron
diffraction technique. Similar to the temperature effect, the block
antiferromagnetic order gradually decreases upon increasing pressure while the Fe vacancy superstructural order
remains intact before its precipitous disappearance at the critical
pressure $P_c=$8.3~GPa.  Combined
with previously determined $P_c$ for superconductivity, our
phase diagram under pressure reveals the concurrence of the block
AFM order, the $\sqrt{5}\times\sqrt{5}$ iron vacancy order and superconductivity for the 245 superconductor. A synthesis of current experimental data in a coherent physical picture is attempted.
\end{abstract}


\maketitle

The recently discovered metal-intercalated iron selenide superconductors
$A_{2}$Fe$_{4}$Se$_5$ ($A$=K, Cs, Tl-K, Rb, Tl-Rb) (245) compounds, with $T_c\sim 30$
K, have attracted much interest \cite{C122924,E106501}.  A high
transition-temperature ($T_N\approx 470$-560 K) and large magnetic moment
(3.3$\mu_B$/Fe) block antiferromagnetic (AFM) order exists in the
superconducting samples \cite{D011873,D020830,D022882}. And magnetic order-parameter experiences an
anomaly when $T_c$ is approached \cite{D020830,D022882}.  The superconductors
crystallize with a highly ordered $\sqrt{5}\times\sqrt{5}$ superstructure, in
which the Fe1 site of the $I4/m$ structure is only a few percent occupied and
the Fe2 site fully occupied \cite{D020830,D014882}.  The non-superconducting
samples at low-$T$ also crystallize in the $I4/m$ structure, but both
Fe sites are {\em fractionally} occupied \cite{D020488,D023674}, since the numbers of the Fe vacancies in the samples and the vacant sites in the $\sqrt{5}\times\sqrt{5}$ pattern are mismatched. The partially
ordered $\sqrt{5}\times\sqrt{5}$ vacancy order becomes one of three competing
phases for temperature below the room temperature up to $\sim 500$ K, namely,
these samples are phase-separated and in the miscibility
gap at ambient condition \cite{D023674,D012059}. 

Close to the miscibility gap, it is not surprising that the nonstoichiometric
245 superconductors often contain several phases of different space-group symmetry.
It has been a complex and controversial issue to determine the sample composition of the superconductors. 
The KFe$_{1.5}$Se$_2$ (234) of the orthorhombic Fe vacancy order has been
proposed as the parent compound \cite{E041316}. However, this phase is not even the
ground state for KFe$_{1.5}$Se$_2$, and a partially ordered $\sqrt{5}\times\sqrt{5}$ vacancy
superlattice is more stable at low temperature \cite{D023674}.
The KFe$_2$Se$_2$ (122) of $I4/mmm$ symmetry has also been proposed as the
superconducting phase \cite{D080069}. But its
existence in films grown by molecular beam epitaxy method likely requires charge transfer with the substrate,
and there is no trace of its existence in bulk superconducting samples \cite{D020830,E091650}.
Detected in the 245 superconductors is the alkaline metal deficient $A_{x}$Fe$_{2}$Se$_2$ ($x\sim 0.3$-0.6)
phase embedded in $\sqrt{5}\times\sqrt{5}$ iron vacancy ordered
superstructure \cite{E031834,E091650,E090383}, forming various microstructure patterns in plane \cite{D070412,E084159} or heterostructure along the $c$-axis \cite{D012059,E025446,E031834} depending on sample preparation procedures. The average sample compositions of these superconductors are consistent with the phase diagram in \cite{D023674}. The question is what role the $A_{x}$Fe$_{2}$Se$_2$ ($x\sim 0.3$-0.6), the $\sqrt{5}\times\sqrt{5}$ superstructure and the AFM order play in the 245 superconductors.

High pressure adds an additional dimension to the complex composition phase-diagram
of 245 superconductors \cite{D023674}, offering a ``clean'' way to investigate
the relation among various phases \cite{D010092,D106138}.  The $T_c$ has been
suppressed to zero at critical pressure $P_c\approx 6$ GPa for $A$= Rb
\cite{D106138,D123822}, 8 GPa for $A$= Cs \cite{D022464}, and 9 GPa for $A$= K
and Tl-Rb superconductors \cite{SunLL12}. In the latter study,
superconductivity of a higher $T_c=48$~K is reported to re-emerge between 11
and 13 GPa \cite{SunLL12}.  High-pressure x-ray powder diffraction experiments
have been performed at room temperature, but differing results have been
reported: the $I4/m$ phase is replaced by an $I4/mmm$ phase at $P_c$ in one
study \cite{D010092}, but the $I4/m$ phase remains up to 15.6 GPa well above
$P_c$ in the other \cite{D123822}.  
In a high-pressure M\"{o}ssbauer spectroscopic study, it has been concluded that
the $A_{x}$Fe$_{2}$Se$_2$ phase in the sample remains intact up to 13.8 GPa.
What has changed is the AFM order on the $\sqrt{5}\times\sqrt{5}$ superstructure which is {\em partially} replaced by a new paramagnetic phase after the superconductivity is suppressed at the critical pressure $P_c$ \cite{D123822}.
Therefore, no clear relationship has been
established between the superconductivity and either the $A_{x}$Fe$_{2}$Se$_2$
phase, the $\sqrt{5}\times\sqrt{5}$ superstructure, or the AFM order in current high
pressure studies.

Here we report high-pressure single-crystal neutron diffraction study of the
(Tl,Rb)$_2$Fe$_4$Se$_5$ superconductor up to 9 GPa, measuring simultaneously the AFM order and the
crystal structure.  The
$\sqrt{5}\times\sqrt{5}$ vacancy order persists under pressure until its
precipitous destruction near $P_c\approx$8.3 GPa when the AFM order parameter is reduced
progressively to zero.  The disappearance of the magnetic and structural
orders coincides with the suppression of superconductivity, revealing the
importance of the block AFM order and the $I4/m$ vacancy order in stabilizing 
superconductivity in the 245 superconductor. 

\begin{figure}[tb!]
\label{fig1}
\includegraphics [width=\columnwidth]{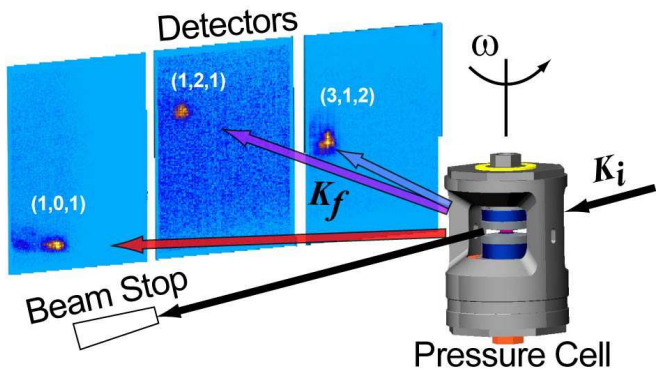}
\caption{(color online)
Schematic diagram of the single-crystal neutron diffraction experiments at
SNAP. The semi-white neutron beam reaches the sample inside the anvil cell and
is diffracted into the position sensitive detectors. 
}
\end{figure}

Single crystals of (Tl,Rb)$_2$Fe$_4$Se$_5$  ($T_c \approx 32.5$ K) were grown
using the Bridgman method \cite{D010462}. They showed sharp diamagnetic
transition at $T_c$ and samples made in the same way had been used in previous
single-crystal neutron scattering studies \cite{D022882,E012413}.  An
orientated crystal of $3\times1\times0.5 mm^3$ was loaded into a
Paris-Edinburgh high-pressure cell together with lead powder serving as the
pressure transmitting medium as well as the pressure gauge \cite{schulte95}.
Neutron diffraction experiments were carried out using the
SNAP instrument at the Spallation Neutron Source (SNS) at the Oak Ridge
National Laboratory (ORNL).  Two separate banks of position sensitive
detectors were centered at the scattering angle $2\theta=48.9^{\circ}$ and
90$^\circ$, respectively. Two wavelength ranges were used during the time-of-flight
diffraction measurements: 1) 0.5 $\leq \lambda \leq 3.8 \AA$ and 2) 4.5 $\leq
\lambda \leq 8.3 \AA$ to access different d-spacing ranges.  The data were
collected up to 9 GPa at 297~K, and 5.7 GPa at 365~K.  Pressure was adjusted
at constant temperature during the measurements.  We label the wavevector
transfer ${\bf Q}=(H,K,L)$ using the tetragonal $I4/m$ unit cell, with $a =
8.683$ and $c=14.39~\AA$ at ambient pressure \cite{D022882}.

Fig.~1 shows the schematic setup of the single crystal experiments. The
superlattice reflections of AFM and Fe-vacancy orders are well separated from
main nuclear Bragg peaks and appear at different regions of the detector
banks.  This is an advantage over our unpublished high-pressure powder diffraction
study, where the magnetic (1,0,1) peak is close to the nuclear (0,0,2)
peak. Their similar $d$-spacings hinder a reliable analysis of the
high-pressure powder neutron diffraction data.  

Fig.~2(a)-(b) show the diffraction peak profile at selected pressures for
the magnetic (1,0,1) and the vacancy superlattice (1,2,1) reflections at 365
K.  Both peaks are smoothly suppressed in intensity without splitting or
appreciable broadening.  Fitting of the integrated intensities of the peaks as
a function of pressure yields critical pressure $P_M(365\rm K)=5.3(2)$~GPa for
the AFM order and $P_S(365\rm K)=5.9(2)$~GPa for the $\sqrt{5}\times\sqrt{5}$
Fe vacancy superstructure.  Fig.~3(a)-3(b) show contour plots of the
diffraction data at 365 K. Clearly the magnetic peak disappears at a lower
pressure than the vacancy superlattice peak.  Figs.~3(d)-(e) present contour
plots for the same two Bragg peaks at 297 K.  Both resolution-limited peaks
indicate that the vacancy and magnetic orders remain long-ranged before their
suppression by high pressure.

\begin{figure}[tb!]
\label{fig2}
\includegraphics[width=\columnwidth]{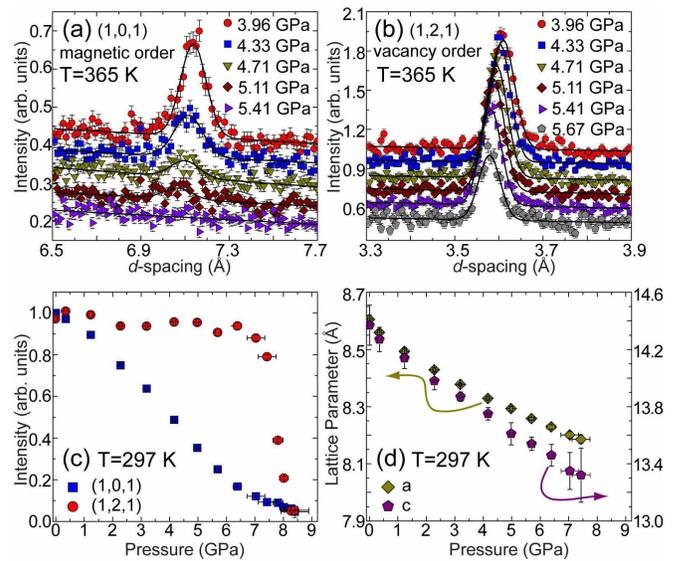}
\caption{(color online) 
(a) Magnetic (1,0,1) and (b) vacancy (1,2,1) peaks at selected pressures at
365 K.  (c) Pressure dependence of the integrated intensity of the vacancy and
magnetic peaks at 297 K.  (d) The lattice parameters of the $I4/m$ structure
as a function of pressure at 297 K. 
}
\end{figure}

In Figs.~3(b) and (e), the (1,2,1) peak of the Fe vacancy
order is suppressed more abruptly than the magnetic (1,0,1) peak in (a) and (d).  Fig.~2(c)
shows the integrated intensity of the two peaks at 297 K respectively.  In
contrast to the gradual suppression of the AFM order, the Fe
vacancy order exhibits a precipitous drop at $P_c \approx8.3$~GPa.  Such
a behavior closely resembles the $T$-dependence of the two long-range
orders at ambient pressure observed in previous neutron diffraction studies
\cite{D020830,D022882}.  Therefore, both at ambient and high pressures, the
order parameter of the $\sqrt{5}\times\sqrt{5}$ vacancy structure reaches the
saturated value when there grows the block AFM order.

Fig.~3(f) shows the nuclear (3,1,2) reflection which survives in the
$I4/mmm$ structure after the suppression of the $\sqrt{5}\times\sqrt{5}$
superstructure under pressure. The lack of peak splitting and absence of
additional reflection in the pressure tuning between the $I4/mmm$ and $I4/m$
structures differ markedly from what have been observed in the temperature tuning
of phase-separated samples \cite{E091650}.  There is an inflection in the peak
position at $P_c$ in Figs.~3(f), indicating lattice parameter relaxation after
the sample experiences the pressure-induced $I4/m$ to $I4/mmm$ structural
transition. The lattice parameters $a$ and $c$ from least-square refinements
from a number of Bragg reflections including (1,0,1), (1,2,1), (3,1,2), and
(5,0,3) at 297 K are shown in Fig.~2(d) as a function of pressure.  Both
shrink smoothly and do not exhibit any anomalies below $P_c$ in the vacancy ordered
state.  Within the Fe vacancy-ordered $I4/m$ phase, (Tl,Rb)$_2$Fe$_4$Se$_5$
exhibits moderate anisotropic compressibility: the lattice parameter $c$ is
reduced by about 9.3\% and the in-plane lattice parameter $a$ decreases by 5\% at
$7.5$~GPa.  This contrasts with the result found in $\rm CaFe_2As_2$, where
the $c$-axis collapses with application of merely 0.4 GPa \cite{A112013} and
the pressure-induced structure transition destroys AFM order without
introducing superconductivity \cite{A112013,A112554}.

\begin{figure}[tb!]
\label{fig3}
\includegraphics[width=\columnwidth]{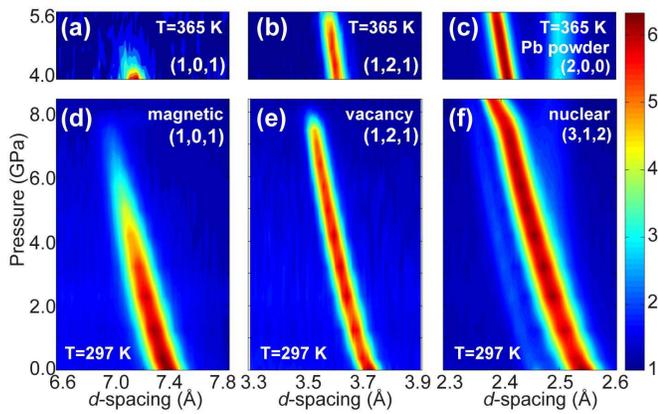}
\caption{(color online)
Contour plots of Bragg intensity of the magnetic peak (1,0,1) at (a) 365 K and
(d) 297 K; the vacancy superlattice peak (1,2,1) at (b) 365 K and (e) 297 K;
(f) the main nuclear Bragg peak (3,1,2) at 297 K. (c) The (200) Bragg peak of
Pb inside the pressure anvil cell at 365 K. The weak pressure-independent
intensity at $d$-spacing 2.51 $\AA$ is the diffraction from the pressure cell.
}
\end{figure}

Sample pressure was monitored {\it in situ} by measuring the $d$-spacing of
the lead (2,0,0) Bragg peak. Fig.~3(c) shows an example of its clear pressure
evolution.  After releasing the pressure from above $P_c$ back to zero at 297 K, all
characteristic reflections associated with the magnetic and vacancy orders in
the $I4/m$ phase reappear. However, intensity of the magnetic peak (1,0,1) is
only 15 percent of the original value at ambient pressure, although intensity
of the vacancy order peak (1,2,1) and the main nuclear peak (3,1,2) are fully
recovered. This indicates that the $\sqrt{5}\times\sqrt{5}$ vacancy order can
sustain the pressure-cycling but the AFM order cannot recover its original
fully ordered state in the constant temperature cycle.

The $P$-$T$ phase diagram based on neutron results is shown in Fig.~4.  At
ambient pressure, Fe vacancies in (Tl,Rb)$_2$Fe$_4$Se$_5$ form a highly
ordered $\sqrt5\times\sqrt5$ superlattice at $T_S\approx 512$~K which is
followed by the block AFM order at $T_N\approx 511$ K.  Both $T_S$ and $T_N$
are suppressed continuously by pressure, and the vacancy and AFM orders are
absent at 9~GPa.  The critical pressure coincides well with the pressure where
the superconductivity was suppressed in the high-pressure resistivity and AC
magnetic susceptibility works on (Tl,Rb)$_2$Fe$_4$Se$_5$ \cite{SunLL12}.  This
phase-diagram provides direct evidence indicating an intimidate connection of
the superconductivity with the block AFM order developing on the
iron-vacancy superlattice in the 245 superconductor.  While $T_S$ and
$T_N$ track each other (Fig. 4), interestingly the superconducting transition
temperature $T_c$ tracks the magnetic (1,0,1) Bragg intensity more closely than
the superlattice (1,2,1) Bragg intensity under pressure.

\begin{figure}[t!]
\label{fig4}
\includegraphics[width=0.93\columnwidth]{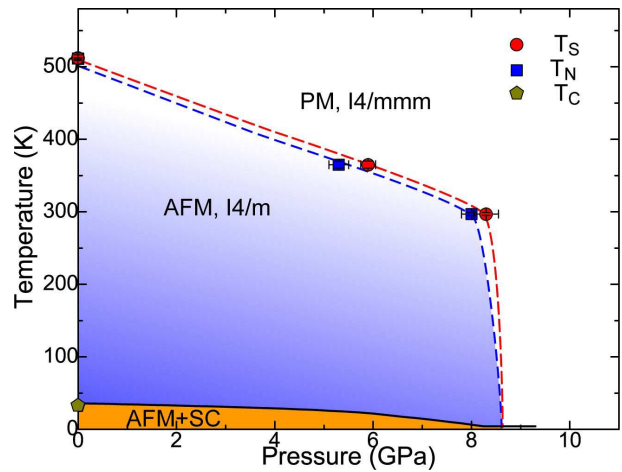}
\caption{(color online)
Pressure-temperature phase diagram of $\rm (Tl,Rb)_2Fe_4Se_5$. Red circles
denote the transition to the Fe $\sqrt5\times\sqrt5$ vacancy order, and blue
squares the block AFM  order. Brown pentagon from magnetization measurement
denotes the superconducting transition at P=0, and $T_c$ at high pressure is
adopted from \cite{SunLL12}. 
}
\end{figure}

Ksenofontov et al.\ reported that both the $A_{x}$Fe$_{2}$Se$_2$ phase and the
vacancy ordered $I4/m$ phase survive well above $P_c$, and
there is no structural transition up to 15.6 GPa \cite{D123822}. The block AFM
order only starts to slowly decrease above $P_c$ in their high-pressure
studies. However, our results clearly show that the
$\sqrt{5}\times\sqrt{5}$ Fe vacancy order and the AFM order are completely suppressed
above $P_c$. The result for the vacancy superstructure order by Guo et al.\ 
is consistent with ours after the poor counting statistics in their
high-pressure {\em powder} x-ray diffraction data is taken into account
\cite{D010092}. Thus, there was likely either an error in the pressure calibration
or too large a pressure gradient in the work of Ksenofontov et al.
Accordingly, their conclusion that the $A_{x}$Fe$_{2}$Se$_2$ phase survives above
$P_c$ may also be in doubt, although
nobody has been able to detect the weak signal from
the $A_{x}$Fe$_{2}$Se$_2$ phase in any high-pressure diffraction experiment so far. 

The importance of the Fe vacancy order 
has also been demonstrated by transport property. The metal-like resistivity behavior
is a precursor to superconductivity and corresponds to a highly ordered $\sqrt{5}\times\sqrt{5}$ Fe vacancy superlattice \cite{D020830,D023674}. The nonsuperconducting samples, on the other hand,
contain imperfect Fe vacancy order which introduces substantial site disorder.
The similar property has been shown in the Fe$_{1+x}$(Se,Te) superconductors for which disordered spin scattering induced by the interstitial excess Fe is responsible for the metal-semiconductor crossover
\cite{C035647}. 
The scanning tunneling spectroscopy
study on KFe$_2$Se$_2$ films, in which random Fe vacancies serve as
spin carrying scatterers, also show the same microscopic behavior that is destructive to the local superconducting gap \cite{D080069}. 

Additionally, the $\sqrt{5}\times\sqrt{5}$ vacancy order with its associated AFM order is substantially more stable with the magnetostructural energy gain through the formation of the
Fe tetramers \cite{D021344,D022215}.
It is thus conceivable that the few percent Fe at the minority Fe1 site
in the average $I4/m$ structure of the 245 superconductors \cite{D020830,D014882,E091650}, instead of
being randomly distributed, aggregates to form nanoscale phase separation in order to save energy in breaking up the tetramers.  Close interaction between the superconducting and AFM order parameters is therefore expected. When the excess Fe at the Fe1 sites
aggregate on the $\sqrt{5}\times\sqrt{5}$ superlattice of fully ordered Fe2
sites, site disorder is minimized and so is the pair-breaking electron
scattering. The local composition inside the aggregation is
$A_{0.8}$Fe$_{2}$Se$_2$ and outside it $A_{0.8}$Fe$_{1.6}$Se$_2$, which
average to a $A_{0.8}$Fe$_{1.6+\delta}$Se$_2$ sample composition. 
When the highly ordered $I4/m$ phase is suppressed at high pressure for the superconducting samples 
or is upset in nonsuperconducting samples, the energetics driving the formation of phase segregation
is lost, so is the superconductivity. 

In summary, we have performed high-pressure single-crystal neutron diffraction
study on magnetic and structural transitions in $\rm (Tl,Rl)_2Fe_4Se_5$
superconductor.  We found both the $\sqrt{5}\times\sqrt{5}$ Fe vacancy order
and the block AFM order are suppressed at $P_c=8.3$~GPa, where
superconductivity also diminishes.  As in previous temperature dependent
studies, the AFM order is also instrumental in the stability of the Fe vacancy
order under pressure.  Our results demonstrate that the highly ordered
$\sqrt{5}\times\sqrt{5}$ vacancy order and associated block AFM order are
crucial ingredients in the realization of 245 superconductors. 

The works at RUC and ZU were supported by the National Basic Research Program
of China Grant Nos.\ 2012CB921700, 2011CBA00112, 2011CBA00103 and
2012CB821404, by the National Science Foundation of China Grant Nos.\
11034012, 11190024, 11374261 and 11204059, and by Zhejiang Provincial Natural
Science Foundation Grant No.\ LQ12A04007.  Research at ORNL's SNS was
sponsored by the Scientific User Facilities Division, Office of Basic Energy
Sciences, U.S.\ DOE.


\begin{thebibliography}{10}

\bibitem{C122924}
J.~Guo {\em et~al.},
\newblock Phys. Rev. B {\bf 82}, 180520(R) (2010).

\bibitem{E106501}
E.~Dagotto,
\newblock Rev. Mod. Phys. {\bf 85}, 849 (2012).

\bibitem{D011873}
Z.~Shermadini {\em et~al.},
\newblock Phys. Rev. Lett. {\bf 106}, 117602 (2011).

\bibitem{D020830}
W.~Bao {\em et~al.},
\newblock Chin. Phys. Lett. {\bf 28}, 086104 (2011).

\bibitem{D022882}
F.~Ye {\em et~al.},
\newblock Phys. Rev. Lett. {\bf 107}, 137003 (2011).

\bibitem{D014882}
P.~Zavalij {\em et~al.},
\newblock Phys. Rev. B {\bf 83}, 132509 (2011).

\bibitem{D020488}
J.~Bacsa {\em et~al.},
\newblock Chem. Sci. {\bf 2}, 1054 (2011).

\bibitem{D023674}
W.~Bao {\em et~al.},
\newblock Chin. Phys. Lett. {\bf 30}, 027402 (2013).

\bibitem{D012059}
Z.~Wang {\em et~al.},
\newblock Phys. Rev. B {\bf 83}, 140505 (2011).

\bibitem{E041316}
J.~Zhao, H.~Cao, E.~Bourret-Courchesne, D.-H. Lee, and R.~J. Birgeneau,
\newblock Phys. Rev. Lett. {\bf 109}, 267003 (2012).

\bibitem{D080069}
W.~Li {\em et~al.},
\newblock Nature Phys. {\bf 8}, 126 (2011).

\bibitem{E091650}
D.~P. Shoemaker {\em et~al.},
\newblock Phys. Rev. B {\bf 86}, 184511 (2012).

\bibitem{E031834}
Y.~Texier {\em et~al.},
\newblock Phys. Rev. Lett. {\bf 108}, 237002 (2012).

\bibitem{E090383}
S.~Weyeneth {\em et~al.},
\newblock Phys. Rev. B {\bf 86}, 134530 (2012).

\bibitem{D070412}
A.~Ricci {\em et~al.},
\newblock Phys. Rev. B {\bf 84}, 060511(R) (2011).

\bibitem{E084159}
Y.~Liu, Q.~Xing, K.~W. Dennis, R.~W. McCallum, and T.~A. Lograsso,
\newblock Phys. Rev. B {\bf 86}, 144507 (2012).

\bibitem{E025446}
A.~Charnukha {\em et~al.},
\newblock Phys. Rev. Lett. {\bf 109}, 017003 (2012).

\bibitem{D010092}
J.~Guo {\em et~al.},
\newblock Phys. Rev. Lett. {\bf 108}, 197001 (2012).

\bibitem{D106138}
M.~Gooch {\em et~al.},
\newblock Phys. Rev. B {\bf 84}, 184517 (2011).

\bibitem{D123822}
V.~Ksenofontov {\em et~al.},
\newblock Phys. Rev. B {\bf 85}, 214519 (2012).

\bibitem{D022464}
G.~Seyfarth {\em et~al.},
\newblock Solid State Communications {\bf 151}, 747 (2011).

\bibitem{SunLL12}
L.~L. Sun {\em et~al.},
\newblock Nature {\bf 483}, 67 (2012).

\bibitem{D010462}
H.~Wang {\em et~al.},
\newblock Europhys. Lett. {\bf 93}, 47004 (2011).

\bibitem{E012413}
S.~Chi {\em et~al.},
\newblock Phys. Rev. B {\bf 87}, 100501(R) (2013).

\bibitem{schulte95}
O.~Schulte and W.~B. Holzapfel,
\newblock Phys. Rev. B {\bf 52}, 12636 (1995).

\bibitem{A112013}
A.~Goldman {\em et~al.},
\newblock Phys. Rev. B {\bf 79}, 024513 (2009).

\bibitem{A112554}
W.~Yu {\em et~al.},
\newblock Phys. Rev. B {\bf 79}, 020511(R) (2009).

\bibitem{C035647}
T.~Liu {\em et~al.},
\newblock Nature Mat. {\bf 9}, 716 (2010).

\bibitem{D021344}
C.~Cao and J.~Dai,
\newblock Phys. Rev. Lett. {\bf 107}, 056401 (2011).

\bibitem{D022215}
X.-W. Yan, M.~Gao, Z.-Y. Lu, and T.~Xiang,
\newblock Phys. Rev. B {\bf 83}, 233205 (2011).

\end{thebibliography}

\end{document}